\documentclass{PoS}
\usepackage{subfigure}
\title{Kaon oscillations in the Standard Model and Beyond using $N_f=2$ dynamical quarks}

\ShortTitle{$K$-bag parameter  from tmQCD}
\author{ETM Collaboration}
\author{V.~Bertone, P.~Dimopoulos, R.~Frezzotti, G.C.~Rossi, A.~Vladikas$^{\dagger}$\\
        Dip. di Fisica, Universit\`a di Roma ``Tor Vergata",\\
        $^{\dagger}$ INFN-``Tor Vergata"\\
        Via della Ricerca Scientifica 1, I-00133 Rome, Italy \\
        E-mail: \email{\{bertone,dimopoulos,frezzotti,rossig,vladikas\}@roma2.infn.it}}

\author{V.~Gimenez, D.~Palao\\
        Dep. de Fisica Te\`orica and IFIC, Univ. de Val\`encia-CSIC,\\
        Dr.~Moliner 50, E-46100 Val\`encia, Spain\\
        E-mail: \email{vicente.gimenez@uv.es, dpalao@roma2.infn.it}}

\author{V.~Lubicz, \speaker{S.~Simula}$^{\ddagger}$\\
         Dip. di Fisica, Universit{\`a} di Roma Tre,\\ 
         $^{\ddagger}$ INFN-Roma Tre \\
         Via della Vasca Navale 84, I-00146 Rome, Italy \\
         E-mail: \email{lubicz@fis.uniroma3.it, simula@roma3.infn.it}}

\author{F.~Mescia\\
        Departament d'Estructura i Constituents de la Mat\`eria and
        Institut de Ci\`encies del Cosmos (ICCUB), Universitat de Barcelona, Diagonal 647, 
        08028 Barcelona, Spain \\      
        E-mail: \email{mescia@ub.edu}}
\author{M.~Papinutto\\
        Laboratoire de Physique Subatomique et de Cosmologie, UJF/CNRS-IN2P3/INPG,\\
        53 rue des Martyrs, 38026 Grenoble, France \\
        E-mail: \email{Mauro.Papinutto@lpsc.in2p3.fr}}

\abstract{
We compute non-perturbatively the B-parameters of the complete basis of
four-fermion operators needed to study the Kaon oscillations in the SM and in its supersymmetric
extension. We perform numerical simulations with
two dynamical maximally twisted sea quarks at three values of the lattice spacing 
on configurations generated by the ETMC. 
Unwanted operator mixings and $O(a)$ discretization effects are removed
by discretizing the valence quarks with a suitable Osterwalder-Seiler variant
of the Twisted Mass action. Operators are renormalized non-perturbatively in
the RI/MOM scheme. Our preliminary result for $B_{{\rm K}}^{{\rm RGI}}$ is 0.73(3)(3).
}

\FullConference{The XXVII International Symposium on Lattice Field Theory--LAT2009\\
		 July 26-31  2009\\
		 Peking University, Beijing, China }

\begin{document}

\section{Introductory Remarks and Calculation Setup}

We will present the main features of the method and preliminary results of
the bag parameter calculation for the K meson oscillations 
at three values of the lattice spacing  
using the $N_f=2$ dynamical 
quark configurations produced by the ETM collaboration. 

ETMC dynamical configurations have been produced with the tree-level Symmanzik
improved action in the gauge sector while the dynamical quarks have been regularized 
by employing the twisted mass (tm) formalism \cite{tmQCD1}. It has been demonstrated that 
with the condition of {\it maximal twist} this formalism provides automatic 
$O(a)$-improved physical
quantities \cite{Frezz-Rossi1}. 

In the so called physical basis the fermion lattice action concerning the sea sector is written
\begin{equation} \label{sea-action}
S_{sea} = a^4 \sum_{x} \bar\psi(x) (\gamma \tilde{\nabla} -i \gamma_5 ~\tau_3~ W_{cr} + \mu_{sea}
) \psi(x) \,\,\, ,
\end{equation}
with $ W_{cr} = -\frac{a}{2} \sum_{\mu} \nabla_{\mu}^{*} \nabla_{\mu} + M_{cr}(r=1)$;
$\psi = (u ~~d)^{T}$ is a doublet of degenerate light sea quarks while
$\mu_{sea}={\rm diag}(\mu_u ~~\mu_d)$.
We should also note that the tm formalism offers a simpler renormalisation pattern 
with comparison to  the standard Wilson regularization. This is true for some  
important physical quantities calculated on the lattice, as for example  the pseudoscalar
decay constant and the chiral condensate.             

It has been shown that the use of the tm regularization can simplify the renormalization pattern 
properties of the four-fermion operators which enter in the calculation of  
certain phenomenologically important weak matrix elements such as $B_K$ 
\cite{tmQCD1, AlphaBK, PenSinVla}. 
In order to achieve both $O(a)$ improvement and a continuum-like
renormalization pattern in the evaluation of $B_K$ we introduce the
valence quarks with Osterwalder-Seiler lattice action and allow for
replica of the down ($d$, $d'$) and strange ($s$, $s'$)
flavours~\cite{Frezz-Rossi2}, viz.\
\begin{equation} \label{action}
S_{val} = a^4 \sum_{x} \sum_{f=d,d',s,s'} \bar{q}_f(x) \,
\Big( \gamma \tilde{\nabla} -i \gamma_5 ~r_f~ W_{cr} + \mu_f  \Big) \, q_f(x)
\; , \qquad -r_s = r_d = r_{d'} =r_{s'} = 1 \, .
\end{equation}
The valence sector action above is written (unlike eq.~(\ref{sea-action})) in the so called
physical quark basis with the field $q_f$ representing just one individual flavour. While the
four fermion operator relevant for $B_K$ (see eq.~(\ref{operator})) is chosen
to contain all the four valence flavours in eq.~(\ref{action}), the interpolating fields for the
external (anti)Kaon states are made up of a tm-quark pair ($\bar{d}\gamma_5 s$,
with $-r_s = r_d$) and a OS-quark pair ($\bar{d}'\gamma_5 s'$, with $r_{d'} =r_{s'}$).
This mixed action setup with maximally twisted Wilson-like quarks has been studied in
detail in Ref.~\cite{Frezz-Rossi2}, allows for an easy matching of sea and valence
quark masses and leads to unitarity violations that vanish as $a^2$ as the continuum
limit is approached. In the present case, however, the quark mass matching is incomplete
because we are neglecting the sea strange quark (i.e. partially quanched computation), 
thereby inducing some
(possibly small) O($a^0$) systematic error.
We notice that the proposed method for obtaining  automatic $O(a)$ improved results 
has already been tested successfully in the calculation of $B_K$ with fully quenched quarks
\cite{ALPHA-BK-2009}. 
 
In Table~\ref{simuldetails} we give the  simulation details 
concerning the mass values of the sea 
and the  valence quarks  for each value of the gauge coupling 
for the calculation presented in this work.
The smallest sea quark mass corresponds to a pion of about 270 MeV 
for  the case of $\beta=3.90$. For $\beta=4.05$ the lightest pion weights 300 MeV while
for $\beta=3.80$ the lowest pion mass is around 400 MeV.
The highest sea quark mass for the three values of the lattice spacing is about
half the strange quark mass.
For the inversions in the valence sector
we have made use of the stochastic method (one--end trick of ref.~\cite{Michael}) in order
to increase the statistical information. Propagator sources have been located  at randomly 
chosen timeslices. For more details on the dynamical configurations and the stochastic 
method application see Refs~\cite{etmc-light, etmc-long}.

\begin{table}[!h]
\begin{center}
\begin{tabular}{cccccccccc}
\hline \hline
 $\beta$  &&  $a^{-4}(L^3 \times T)$ && $a\mu_{\ell}~=~a\mu_{sea}$      &&  $a\mu_{h}$ &&   \\
\hline
3.80      &&  $24^3 \times 48$&& 0.0080 0.0110   && 0.0200, 0.0250   &&     \\
($a\sim0.1~\mbox{fm}$)          &&                  &&                 && 0.0300, 0.0360   &&     \\
\hline
3.90      &&  $24^3 \times 48$&& 0.0040, 0.0064 &&   0.0150, 0.0220 &&    \\
          &&                  && 0.0085, 0.0100 &&   0.0270, 0.0320         &&    \\
3.90      &&  $32^3 \times 64$&& 0.0030, 0.0040 &&   0.0220, 0.0270 &&    \\
($a\sim0.085~\mbox{fm}$) && && && && \\
\hline
4.05      &&  $32^3 \times 64$&& 0.0030, 0.0060 &&   0.0150, 0.0180 &&      \\
($a\sim0.065~\mbox{fm}$)          &&                  && \hspace*{-1.3cm} 0.0080         &&   0.0220, 0.0260 &&      \\
\hline \hline
\end{tabular}
\end{center}
\caption{Simulation details}
\label{simuldetails}
\end{table}

\section{The K-meson bag parameter}

We recall that in our mixed action setup all the physical quantities are evaluated
with no O($a$) discretization effects (see Ref.~\cite{Frezz-Rossi2}) 
and moreover the four fermion operator relevant for $B_K$, which reads
\begin{equation}
\Big{[} V_\mu V_\mu + A_\mu A_\mu \,\,\Big{]}_{\mbox{bare}}^{\mbox{phys-basis}} =
[(\bar{q}_{s}\gamma_\mu q_d) (\bar{q}_{s'}\gamma_\mu q_{d'}) + (\bar{q}_{s}\gamma_\mu\gamma_5 q_d) (\bar{q}_{s'}\gamma_\mu \gamma_5 q_{d'})]
+ [ d \leftrightarrow d' ] \, .
\label{operator}
\end{equation}
is multiplicatively renormalizable. This can be easily understood by noting that in the (unphysical) tm basis, where the Wilson term enters the valence action in
the standard way (with no $i\gamma_5$-twist) and the operator renormalization
properties are the same of the standard Wilson fermionic action, 
the operator~(\ref{operator}) takes
the form
\begin{equation}
\Big{[} V_\mu A_\mu + A_\mu V_\mu \,\,\Big{]}_{\mbox{bare}}^{\mbox{tm-basis}} =
[(\bar{\chi}_{s}\gamma_\mu \chi_{d}) (\bar{\chi}_{s'}\gamma_\mu \gamma_5 \chi_{d'}) + (\bar{\chi}_{s}\gamma_\mu\gamma_5 \chi_{d}) (\bar{\chi}_{s'}\gamma_\mu \chi_{d'})]
+ [ d \leftrightarrow d' ] \, ,
\label{operator_tmba}
\end{equation}
Here $\chi_f = \exp^{-i\gamma_5 \pi/4} q_f$ and $\bar\chi_f = \bar{q}_f \exp^{-i\gamma_5 \pi/4}$,$f=d,d',s,s'$ are the tm basis valence quark fields.
The operator~(\ref{operator_tmba}) is known
to be protected from mixing under renormalisation due to $CPS$ symmetry~\cite{Bernard}.
In summary we have (``R'' stands for ``renormalized'')
\begin{equation}
\Big{[} V_\mu V_\mu + A_\mu A_\mu \,\,\Big{]}_{\mbox{R}}^{\mbox{phys-basis}} =
\,\, Z_{VA+AV} \,\,
\Big{[} V_\mu V_\mu + A_\mu A_\mu \,\,\Big{]}_{\mbox{bare}}^{\mbox{phys-basis}} =
\,\, Z_{VA+AV} \,\,
\Big{[} V_\mu A_\mu + A_\mu V_\mu \,\,\Big{]}_{\mbox{bare}}^{\mbox{tm-basis}} \, ,
\end{equation}
where the name of the renormalization constant is chosen so as to be consistent
with the notation used in the standard Wilson fermion literature.

In order to estimate the $B_{\rm K}$-parameter we calculate a three-point correlation function 
where a four-fermion operator is free to move in lattice time $t$ whereas 
two ``K-meson walls" consisting of noisy sources are imposed at fixed time separation 
$t_R-t_L = T/2$. The $t_L$ value changes 
randomly from configuration to configuration. 
In our simulations we consider the time reversed case too and we average them properly.     
The plateau signal is taken for $t_L \ll t \ll t_R$. 
We extract $B_{\rm K}$ from the ratio:
\begin{equation}
R_{{\rm B_K}} = \frac{ C^{(3)}_{\bar{K}OK}(t-t_L,t-t_R) }{ C^{(2)}_{\bar{K}}(t-t_L)
C^{(2)}_{K}(t-t_R) } \stackrel{t_L \ll t \ll t_R}\longrightarrow B_{{\rm K}}
\end{equation}
In our analysis  all correlation functions 
satisfy the condition $a\mu_l = a\mu_{\rm sea}$ while the valence strange-like quark mass values 
are given in Table~\ref{simuldetails}. 
An important remark is in order: the mixed regularization set-up that we have used  
leads at finite lattice spacing  
to different values for the decay constant and the pseudoscalar masses 
of the two K-mesons employed in the calculation. 
We find that the discretisation effects are negligible for the 
decay constant while happen to be significant in the case of the pseudoscalar mass. 
For this reason we normalize the four fermion matrix element by dividing with 
$(8/3) m_K^{OS}~ m_K^{tm}~f_K^{OS}~f_K^{tm}$. Moreover, as expected, the cutoff effects 
 diminish drastically towards the CL.  
So this kind of systematic error  is well under control.

The fits to the light quark mass behaviour
are performed using the $SU(2)$ Partially Quenched Chiral Perturbation Theory  formula 
of refs~\cite{SharpeZhang,Alltonetal}. In our case the fit ansatz is:
\begin{equation}
B(\mu_h) \,\, = \,\, B_\chi(\mu_h) \,
\Big [ 1 \, + \,  b(\mu_h) \, \frac{2B_0}{f^2} \, \mu_l \, - \, \frac{2B_0}{32\pi^2 f^2} \mu_l \,
\ln\big (\frac{2B_0\mu_l}{\Lambda_\chi^2} \big ) \Big ] + D(\mu_h) a^2
\label{eq:pqchipt}
\end{equation}
where $\mu_h$ denotes the quark mass values around the 
strange quark (see Table~\ref{simuldetails}).
Thus, the fit procedure consists of a combined fit of  chiral and continuum
extrapolation. We find that the cutoff effects on our data 
are well described by a $\mu_l$-independent (but $\mu_h$-dependent) $O(a^2)$ term.  

Two methods of analysis have been followed. 
The first method relies on  using the information 
for the physical mass values of the up/down and  strange quarks in the continuum limit, 
as they have been estimated in a recent ETMC computation \cite{ETMC-prep1}. 
Note that the implementation of this method 
requires the knowlegde of the quark mass renormalization constant \cite{REN}.  
The second method consists of employing the pseudoscalar masses instead of the quark masses. 
In this case we choose a set of three values of reference pseudoscalar masses made out of 
two strange-like quarks, $M_{hh}$; 
keeping each of them fixed we perform the chiral fits in terms of the light pseudoscalar mass. 
In the end of the procedure 
we estimate $B_{\rm K}$ via an interpolation at the physical point defined by the formula 
$M_{ss}^2 = 2 M_K^2 - M_{\pi}^2 $. 
Both methods give compatible final results within less than one standard 
deviation.        

\begin{figure}[!h]
\begin{center}
\subfigure[]{\includegraphics[scale=0.64,angle=-0]{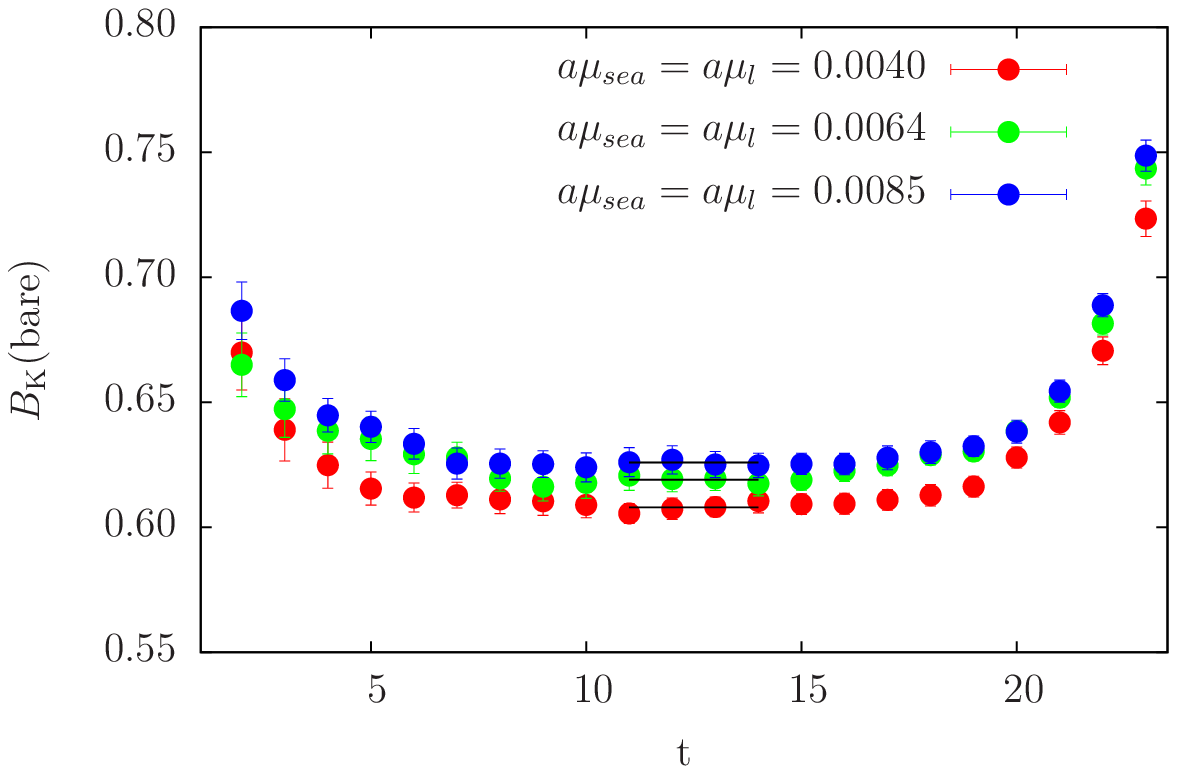}}
\subfigure[]{\includegraphics[scale=0.58, angle=-0]{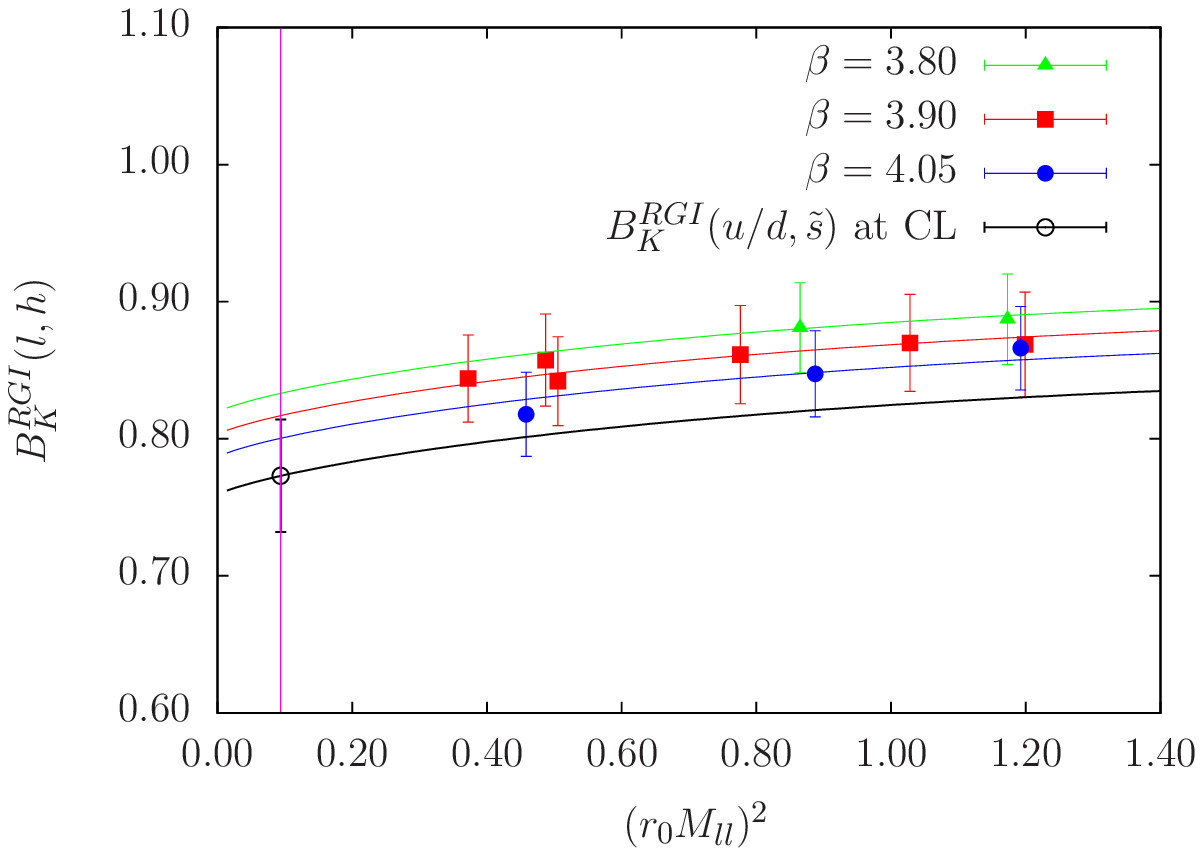}}
\caption[]{(a) The quality of the plateau for three values of the light quark mass for 
$\beta=3.90$; (b) Combined chiral and continuum fit for $B_{\rm K}^{{\rm RGI}}(l,h)$ versus 
$(r_0 M_{ll})^2$. The  
empty black circle gives the value at the continuum limit for the case of $(r0M_{hh})=1.50$; 
}
\label{fig:BK_Z}
\end{center}
\end{figure}

In Figure \ref{fig:BK_Z}(a) the quality of the plateau is shown for $\beta=3.90$,  for
three values of the light quark mass and for one typical value of $\mu_h$;
in Figure \ref{fig:BK_Z}(b) we present an example of a combined chiral plus continuum fit 
(three value of the lattice spacing) for $B_{\rm K}^{{\rm RGI}}(l,h)$ versus the 
light pseudoscalar 
mass squared in units of $r_0$; the value of $M_{hh}$ is in the vicinity of the physical one.

\begin{figure}[!h]
\begin{center}
\subfigure[]{\includegraphics[scale=0.24,angle=-90]{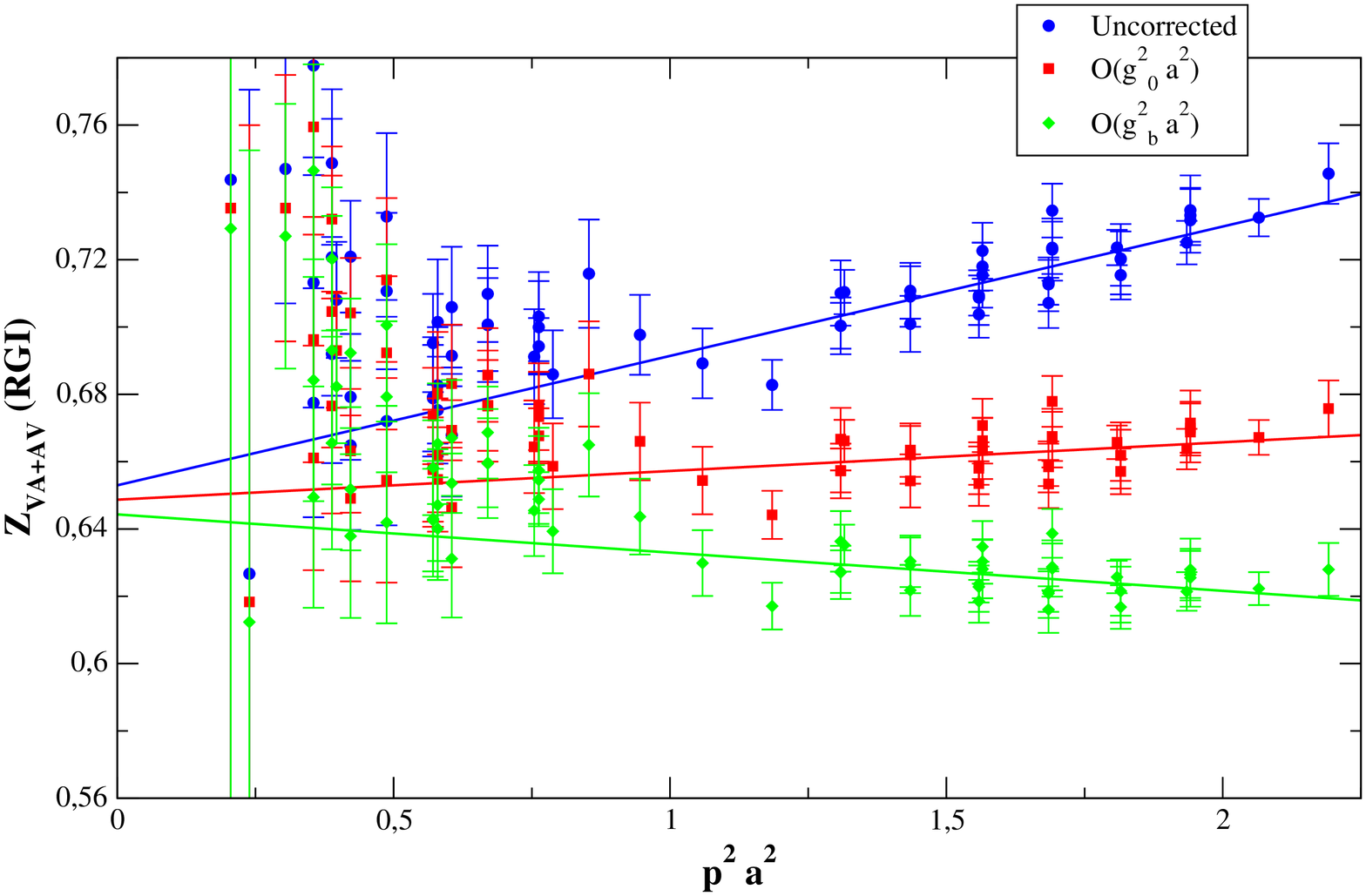}}
\subfigure[]{\includegraphics[scale=0.24,angle=-90]{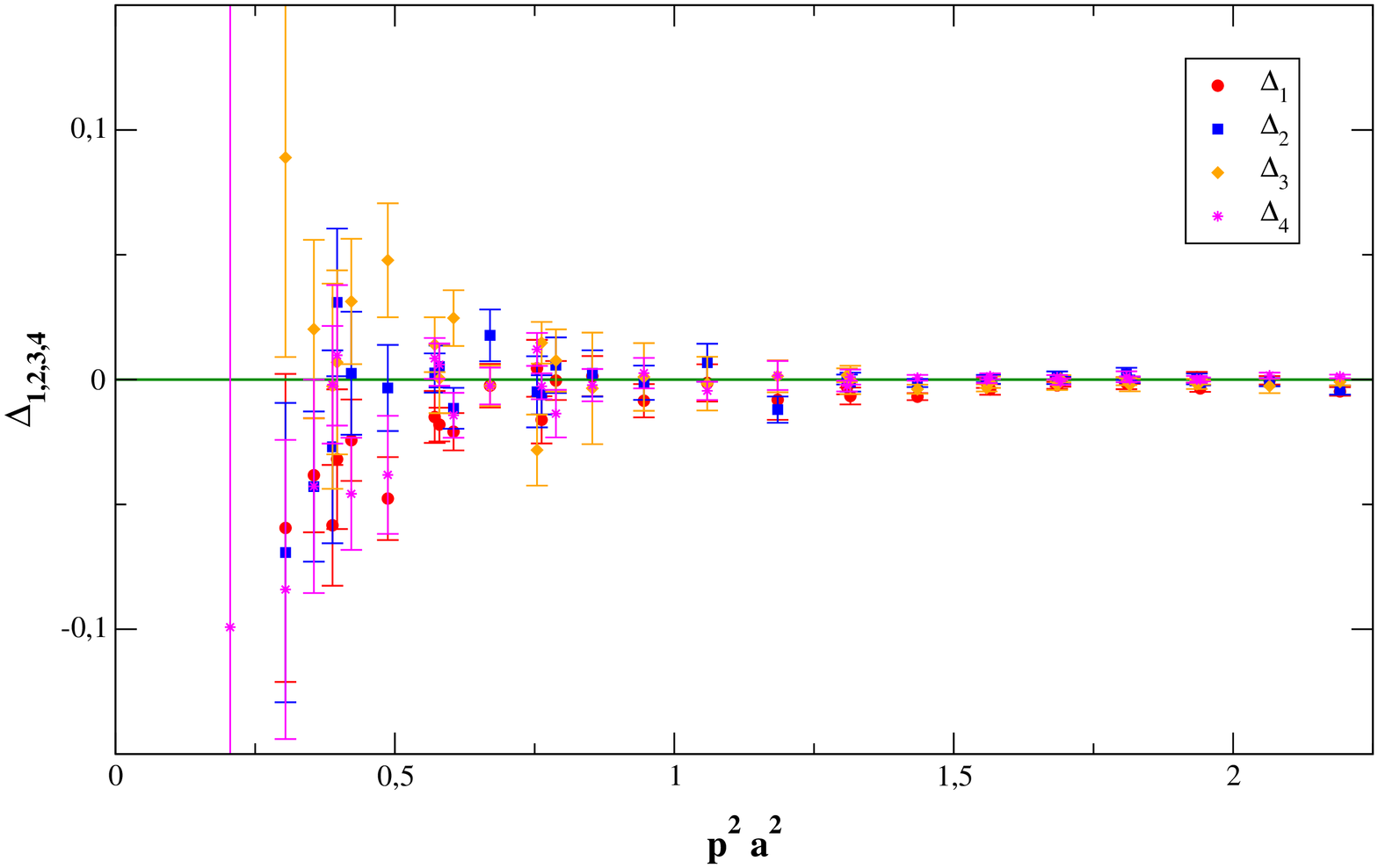}}
\caption[]{(a) RI/MOM computation of the multiplicative renormalization factor 
$Z_{VA+AV}({\rm RGI})$ at $\beta=3.90$;  
(b) Mixing coefficients $\Delta_k$ ($k=1,\cdots ,4$) 
with other four-fermion operators with ``wrong chirality".
}
\label{fig:Zs}
\end{center}
\end{figure}

The two point renormalisation constants for the axial and vector current 
have been calculated using the RI-MOM method \cite{rimom2}. We recall that the physical axial
current made up of OS quarks is normalized by $Z_A$ while the one consisting of 
tm quarks is normalized by $Z_V$ \cite{REN}. 
The RI-MOM method has also been employed for the calculation of the renormalisation constant
of the four-fermion operator \cite{rimom4}. 
In Figure \ref{fig:Zs}(a) we show the behaviour of the renormalisation
constant as a function of the momentum squared in lattice units $(ap)^2$ 
for $\beta=3.90$ at the valence chiral limit and for $a\mu_{sea}=0.0040$. 
Discretization effects of $O(a^2)$ have been evaluated at one loop \cite{Cyp} and
subtracted from the relevant correlation functions. Thus, the leading discretization
effects on our RI-MOM determination of the renormalization constant are of
$O(g^4 a^2,g^2 a^4)$. 
We show three types of results; two of them correspond  to two 
estimates of the subtracted perturbative contributions. 
The amount of the subtraction depends on the choice of the value for the 
gauge coupling. We have considered two cases for the gauge coupling, 
the naive ($g_0$) and the boosted one ($g_b$).
We also show the result for the $Z_{VA+AV}({\rm RGI})$ without considering any 
perturbative subtractions (indicated as ``uncorrected'' in the figure).
In the right panel of the same figure we illustrate  the absence of mixing 
with ``wrong chirality" operators; in fact 
the mixing coefficients are vanishing.  

Our preliminary result for $B_{{\rm K}}$ in the RGI scheme in the continuum limit is 
$$B_{{\rm K}}^{{\rm RGI}}=0.73(3)(3)$$
 The first error includes the  uncertainty 
coming from the  correlators and from the fit procedure (chiral plus continuum) 
while the second one is due to the uncertainties in the
calculation of the renormalisation constants. We are currently attempting to reduce the 
latter uncertainty.

\section{The K-bag meson parameter beyond the SM}

Interactions beyond the SM including supersymmetry furnish new diagrams in the calculation 
of the $\Delta S=2$ process. 
Their effect expressed in the OPE expansion is to enrich the set of the local operators to be
considered in the low energy regime, see e.g. \cite{Ciuchini-etal98}. 
Therefore one has to calculate on the lattice the matrix elements  of five parity even 
four-fermion operators namely $O_1=O_{VV+AA}, O_2=O_{SS+PP}, O_3=O_{\tilde{T}T}, O_4=O_{SS-PP}$ and 
$O_5=O_{VV-AA}$ 
\cite{Allton_etal98, Babich_etal06, Nakamura_etal06}.

It is well known that the renormalisation pattern of the parity-even four-fermion operators 
becomes  complicated because of mixings as soon as  the regularization breaks 
the chiral symmetry; this is certainly the case of  Wilson fermions. 
However using the proposal of 
Ref.~\cite{Frezz-Rossi2} this problem  is bypassed; due to the axial rotation 
mapping of the parity-even to parity-odd operators in the tm basis 
the renormalisation pattern becomes continuum-like  \cite{rimom4}.  
It is worth  mentioning that, as in the case of the SM four-fermion operator, the 
lattice estimates of the matrix elements of $O_2, \ldots, O_5$ are automatically $O(a)$-improved.

\begin{figure}[!h]
\begin{center}
\subfigure[]{\includegraphics[scale=0.29,angle=-90]{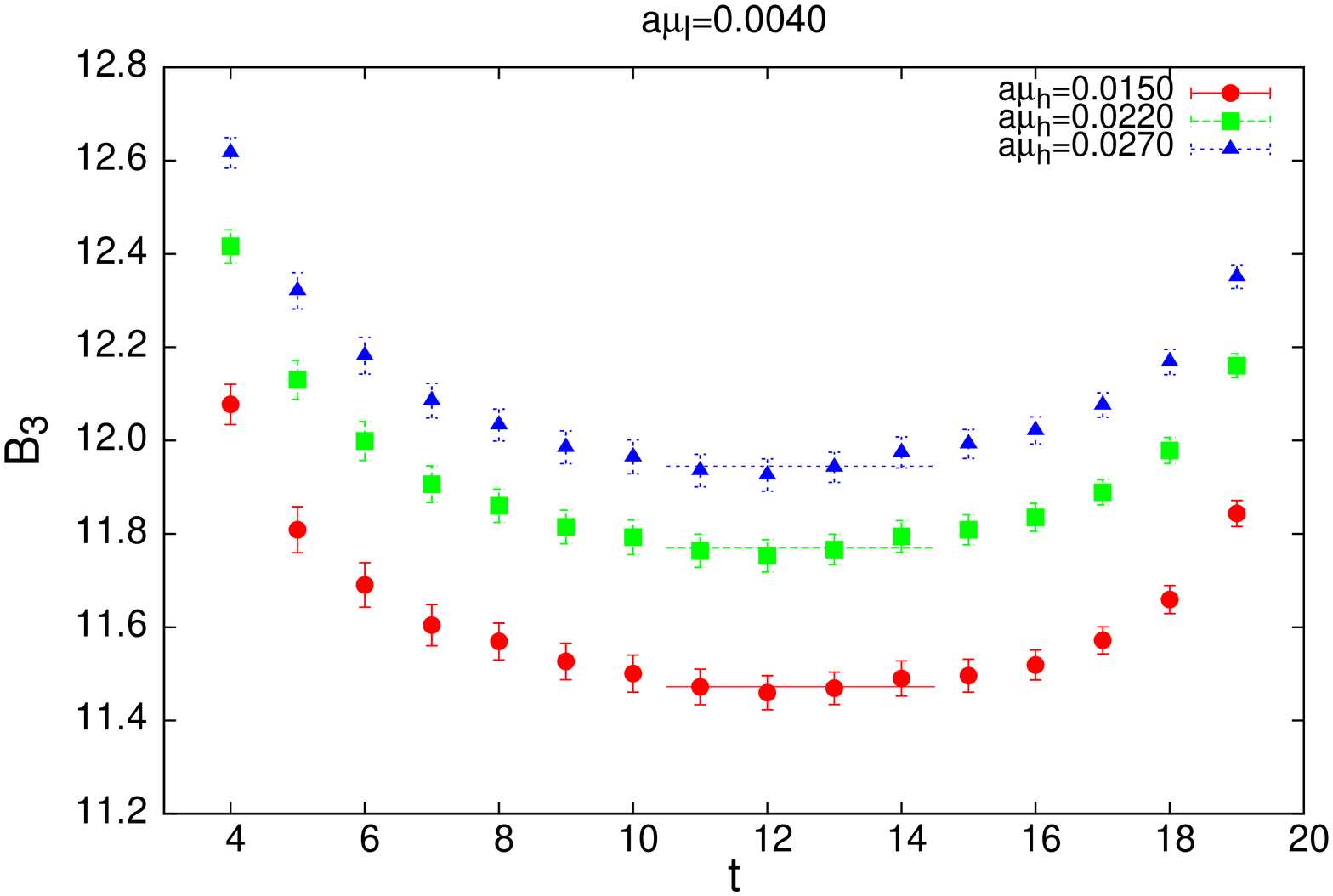}}
\subfigure[]{\includegraphics[scale=0.29,angle=-90]{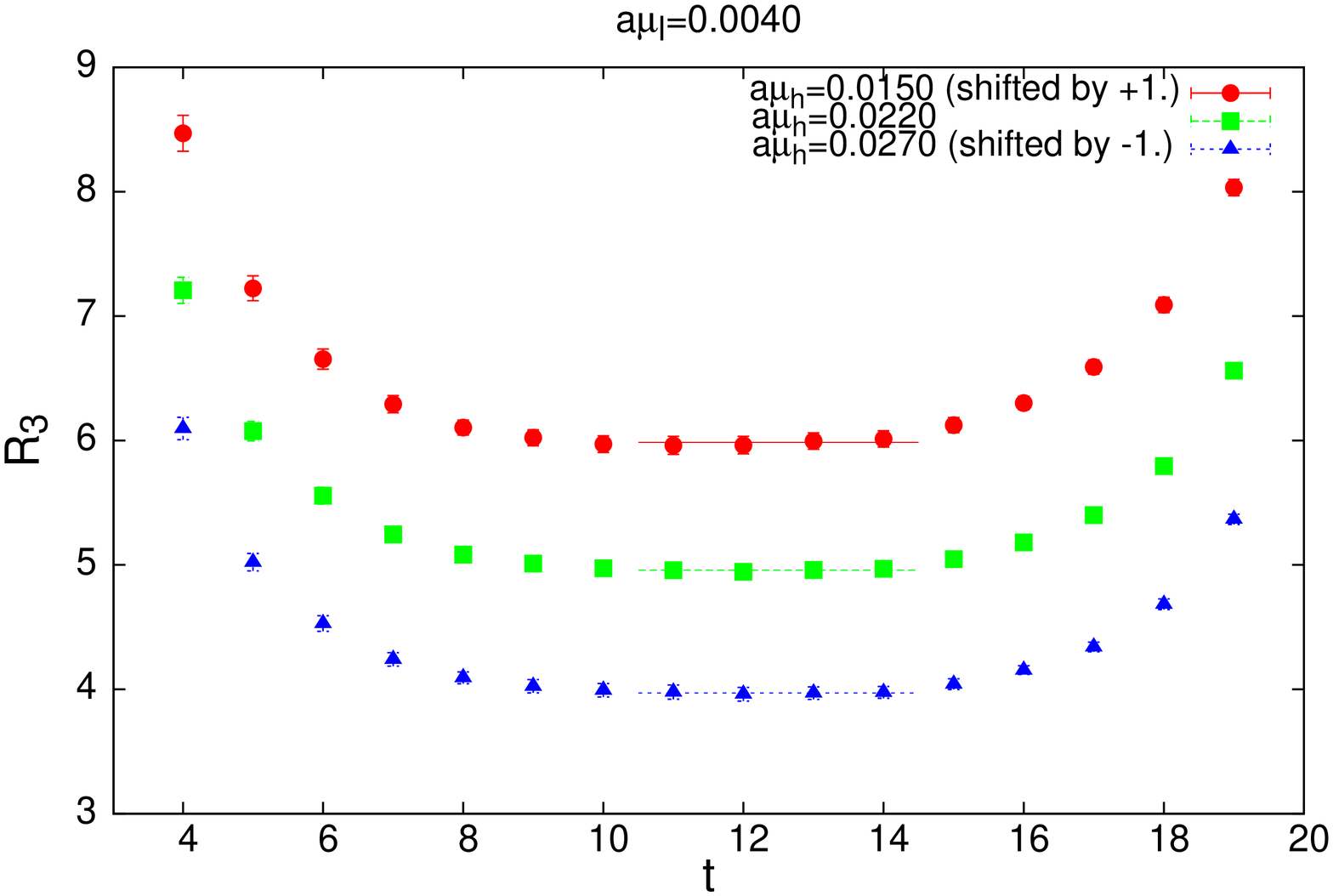}}
\caption[]{(a) and (b): the quality of the signal for the quantities $B_3$ and $R_3$ 
respectively at three values of the strange-like  quark mass using $a\mu_l=0.0040$ 
at $\beta=3.90$.
}
\label{R_SSM}
\end{center}
\end{figure}

First results regarding the signal quality for the case of $\beta=3.90$ are given 
in Figure \ref{R_SSM}. 
We depict the plateaux for the $B_3$ bag parameter (left panel) and  
for the quantity  
$ R_3 \sim \frac{\langle \bar{K}| O_3|K \rangle}{\langle \bar{K}|O_1|K \rangle}  $ (right panel).
Both figures refer to the same value of the light quark mass for three different choices 
of the strange-like quark mass.
Computation at the other two values of the lattice spacing as well as a full determination 
of the renormalisation constant matrix is still in progress.

\section*{Acknowledgements}
We thank our ETM collaborators for their help and encouragement. 
This work has been supported in part by the EU ITN contract MRTN-CT-2006-035482, ``FLAVIAnet''.
F.M. acknowledges the support by CUR Generalitat de Catalunya under project 2009SGR502 and by
the Consolider-Ingenio 2010 Program CPAN (CSD2007-00042), {\it UB-ECM-PF 09/26, ICCUB-09-230}.
V.G. and D.P. thank MICINN (Spain) for partial financial support under grant FPA2008-03373.

\end{document}